\documentclass[sigconf,nonacm]{acmart}
\usepackage{array}
\usepackage{placeins}

\renewcommand\footnotetextcopyrightpermission[1]{}

\title{Exact Adaptive Hybrid Retrieval Without Fixed Top-L Cutoffs}

\author{Chunran Zhang}
\email{chronis@my.swjtu.edu.cn}
\affiliation{%
  \institution{Southwest Jiaotong University}
  \department{School of Computing and Artificial Intelligence}
  \city{Chengdu}
  \country{China}
}

\begin{document}

\begin{abstract}
Modern retrieval-augmented generation (RAG) systems often fuse fixed Top-\(L\)
results from dense and sparse retrievers, treating later contributions as zero.
The cutoff therefore determines both the ranking and its execution cost. Yet
truncated fusion is not generally equivalent to complete-list fusion: unread
cross-list ranks can change Top-\(K\) membership or order even when the observed
candidates contain every item in the complete-list Top-\(K\). Because channel
rankings vary across queries and corpus updates, a depth selected from
historical queries may not transfer reliably.

We propose Exact Adaptive Hybrid Retrieval (EAHR), which fixes the ordered
Top-\(K\) defined by complete-list weighted RRF as the retrieval target and treats
channel depth as request-specific execution state. Per-Vector Scalar
Quantization (PVS) and Posting Block-Max (PBM) produce resumable exact dense and
sparse rankings. Fusion bounds unread contributions and requests further ranks
only while they can change the Top-\(K\). Every successful request therefore
matches complete-list fusion without a preset Top-\(L\); otherwise, execution
continues safely to list exhaustion.

Across five test collections and five temporal corpus snapshots, complete-list
weighted RRF remained competitive, whereas fixed depths selected from
historical queries did not transfer reliably. EAHR reproduced the complete-list
ordered Top-20 in all 150 query--snapshot combinations. Under a warm-cache,
interleaved, order-balanced protocol, the paired geometric-mean latency ratios
of exhaustive batch execution to EAHR were 23.35 on TREC-DL 2019 and 30.28 on
TREC-DL 2020. Anti-correlated rankings exhausted both lists, and some difficult
queries were slower with EAHR. EAHR does not guarantee a speedup for every
request; it fixes the exact result while adapting execution depth to the current
rankings.
\end{abstract}

\ccsdesc[500]{Information systems~Retrieval models and ranking}
\ccsdesc[300]{Information systems~Top-k retrieval in databases}

\keywords{retrieval-augmented generation, hybrid retrieval, rank fusion, exact top-k query processing, vector databases}

\maketitle

\section{Introduction}\label{introduction}

\subsection{\texorpdfstring{Fixed Top-\(L\) fusion: result semantics and depth selection}{Fixed Top-L fusion: result semantics and depth selection}}\label{fixed-top-l-fusion-result-semantics-and-depth-selection}

Hybrid text retrieval commonly combines lexical and semantic retrievers over the same collection \citep{bruch2024fusion, louis2025know}. Their raw scores require additional modeling or calibration before they can be compared \citep{manmatha2001score, louis2025know}. Reciprocal rank fusion (RRF) instead fuses ranks rather than scores; we use its weighted form \citep{cormack2009rrf}.

A common design applies RRF to a fixed Top-\(L\) list from each retriever. Elasticsearch's \path{rank_window_size} and Azure AI Search's finite text and vector result sets exemplify this design \citep{elastic_rrf_docs, azure_rrf_docs}.

Fixed Top-\(L\) truncates each ranked list and treats all later contributions as zero, so it computes weighted RRF over truncated ranked lists instead of complete ones. Even exact Top-\(L\) prefixes do not eliminate these missing contributions: a document observed in one list may still appear at an unread rank in another. These unknown cross-list ranks can change Top-\(K\) membership or order even when the candidate union already contains every document in the Top-\(K\) obtained from complete-list fusion. A fixed prefix of length \(L\) therefore cannot certify the complete-list fusion result. This problem is distinct from ANN approximation error.

Because \(L\) controls both the fusion input and execution cost, it becomes a quality--cost hyperparameter tied to a query sample and corpus snapshot. New queries can produce different dense and sparse rankings on the same corpus; corpus updates can change those rankings for the same query. Open-ended RAG workloads and evolving knowledge bases therefore make a single \(L\) difficult to transfer across requests and snapshots.

\subsection{EAHR: fixed result semantics with adaptive access depth}\label{eahr-fixed-result-semantics-with-adaptive-access-depth}

To return the complete-list fusion result without generating both complete rankings for every query, we propose Exact Adaptive Hybrid Retrieval (EAHR). EAHR fixes this exact ordered Top-\(K\) as the retrieval target and determines each channel's required access depth during execution. EAHR therefore requires no preset per-channel Top-\(L\) cutoff.

\begin{figure*}[t]
\centering
\includegraphics[width=0.75\textwidth]{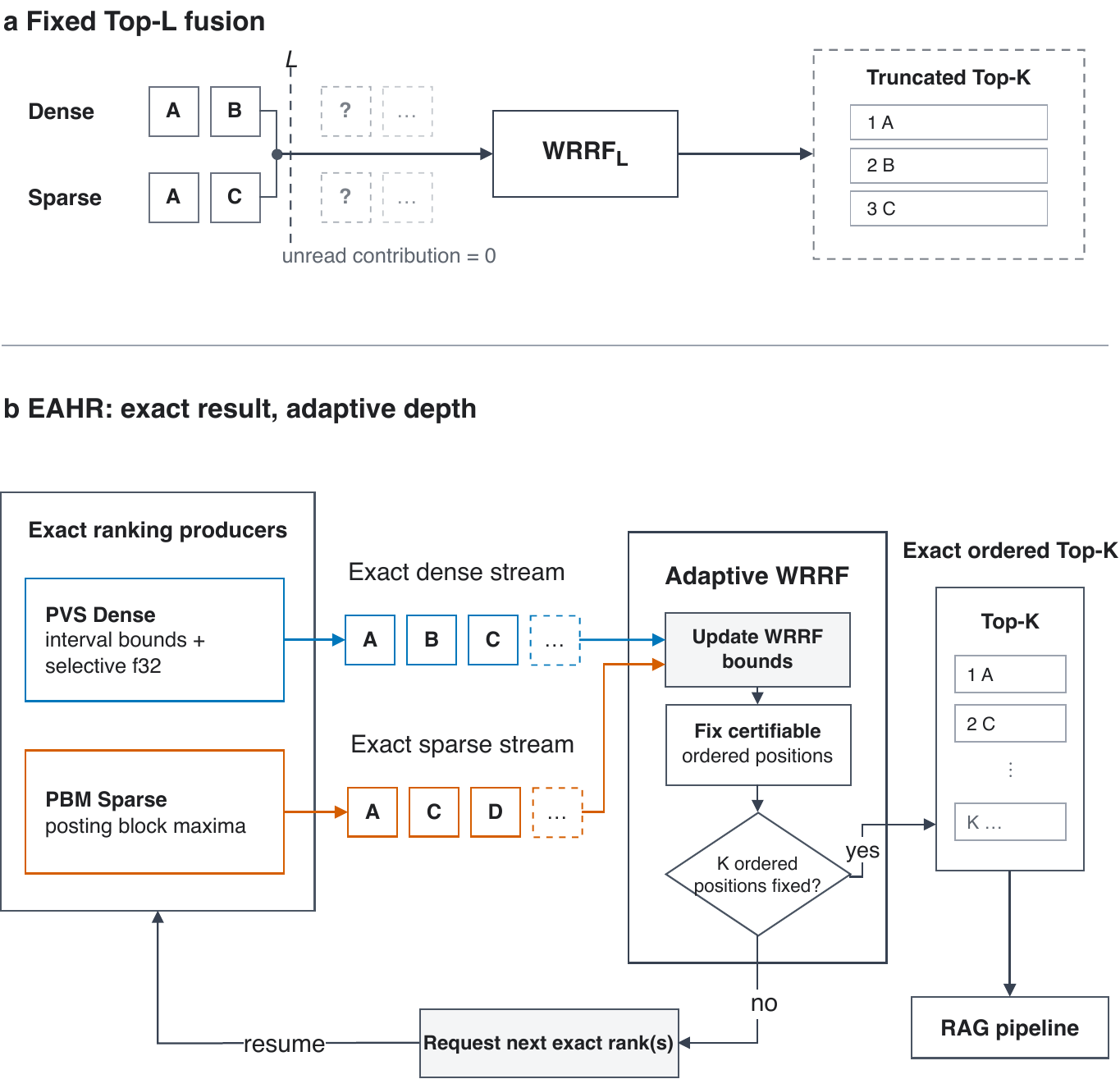}
\caption{Fixed Top-\(L\) defines fusion over truncated rankings. EAHR incrementally produces exact dense and sparse ranks with PVS and PBM until the ordered Top-\(K\) under complete-list weighted RRF is determined.}
\Description{A comparison of fixed Top-L fusion and EAHR. The fixed plan truncates dense and sparse ranked lists before fusion. EAHR incrementally obtains exact rankings from PVS and PBM and continues until the complete-list Top-K is fixed.}
\label{fig:eahr-overview}
\end{figure*}

At the start of a query, neither retriever has scored and ordered every eligible item. EAHR obtains the next exact dense or sparse result only when fusion needs it, while retaining the search state for later continuation. Per-Vector Scalar Quantization (PVS) uses deterministic per-vector score intervals and selective float32 evaluation to determine successive dense results; Posting Block-Max (PBM) uses query-weighted bounds over posting ranges and expands the most promising ranges first to determine sparse results. Both release an item only after ruling out every unprocessed competitor.

Weighted RRF consumes these ranks, bounds contributions from unread positions, and requests more while unresolved items can change Top-\(K\) membership or order. A fixed request snapshot, hierarchical merging, and deterministic tie-breaking keep all batches aligned with one complete-list target. Difficult queries may exhaust a channel or run slower than batch execution; successful responses remain identical in items and order to complete-list fusion.

\subsection{Technical foundations and contributions}\label{technical-foundations-and-contributions}

Recent work retrieves hybrid dense--sparse vectors using a single approximate MIPS score \citep{bruch2024bridging}. This differs from our setting: EAHR keeps the two rankings separate and exactly returns the Top-\(K\) defined by complete-list rank fusion. It must coordinate two operations within the same request: producing the next exact dense or sparse rank and deciding whether the ranks already observed fix the fused order. TA and NRA establish bound-based stopping when sorted lists can be accessed progressively, while MEDRANK applies related ideas to rank aggregation \citep{fagin2003optimal, fagin2003medrank}. Incremental nearest-neighbor search, exact MIPS, and upper-bound pruning for sparse retrieval provide foundations for producing results within individual retrievers \citep{hjaltason1999distance, teflioudi2016lemp, li2017fexipro, abuzaid2019exactmips, broder2003wand, ding2011bmw, mallia2024bmp}. The remaining problem is to coordinate exact rank generation with fusion within one request.

For weighted RRF, we use these bounds to determine when the ordered Top-\(K\) is fixed. To provide the required ranks, we develop PVS and PBM as stateful exact rank generators for dense vectors and compressed sparse postings. A Qdrant query path preserves this behavior across fixed snapshots, hierarchical Segment and Shard merging, deterministic tie-breaking, exhaustion, and failure. Rather than waiting for two fixed Top-\(L\) lists, fusion requests another exact rank only when the current prefixes do not yet fix the result.

Across five test collections, complete-list weighted RRF remained competitive, while the best tested fixed depth varied by collection. In five temporal TREC-COVID snapshots, fixed depths selected from Round 1 did not transfer reliably. EAHR reproduced the complete-list ordered Top-20 in all 150 combinations of queries and snapshots, with access depth varying across queries and changing as the corpus evolved. Under the reported protocol, the paired geometric-mean latency ratios of exhaustive batch execution to EAHR were 23.35 and 30.28 on TREC-DL 2019 and 2020. Anti-correlated rankings exhausted both lists, and some difficult queries were slower with EAHR.

\section{Method}\label{method}

EAHR removes fixed Top-\(L\) from the retrieval request. Given a query, \(K\), a corpus snapshot, and a fusion configuration, it returns the ordered Top-\(K\) defined by complete-list fusion while treating each channel's depth as internal, request-specific execution state.

\subsection{Complete-list fusion target and EAHR execution contract}\label{complete-list-fusion-target-and-eahr-execution-contract}

Let \(U\) be the finite set of items visible to a request under a fixed corpus snapshot, \(\prec_{\mathrm{id}}\) a total order over item identifiers, and \(\mathcal C\) the set of retrievers. Retriever \(i\) ranks a support set \(U_i\subseteq U\); items outside \(U_i\) have no rank and contribute zero to fusion. In our implementation, \(U_i\) contains items with the target vector for dense retrieval and items with positive query scores for sparse retrieval.

Together with \(\prec_{\mathrm{id}}\), the scoring rule of retriever \(i\) defines a unique complete ranked list

\[
\pi_i=(\pi_i(1),\ldots,\pi_i(n_i)),
\qquad n_i=|U_i|,
\]

where \(\pi_i(r)\) is the item at rank \(r\), and \(r_i(x)\) is the rank of \(x\in U_i\).

Let \(g_i(r)\) be the contribution of rank \(r\) from retriever \(i\). We require \(g_i\) to be nonnegative and nonincreasing. The complete-list fusion score of item \(x\) is

\[
F(x)=\sum_{i:x\in U_i}g_i(r_i(x)).
\]

Our Qdrant-based implementation uses the following rank-scaled parameterization of weighted RRF:

\[
g_i(r)=
\begin{cases}
0, & w_i=0,\\[3pt]
\displaystyle\frac{1}{r/w_i+k-1}, & w_i>0,
\end{cases}
\]

Here, \(k\ge 1\) is the rank constant and \(w_i\ge 0\) scales the effective rank of retriever \(i\). This rank-scaled form differs from formulations that place \(w_i\) in the numerator. Because all reported experiments use \(w_i=1\), it reduces to \(g_i(r)=1/(r+k-1)\).

Define

\[
U^+=\{x\in U:F(x)>0\},
\qquad q_K=\min(K,|U^+|).
\]

Sort \(U^+\) by decreasing \(F(x)\), breaking ties according to \(\prec_{\mathrm{id}}\), and let \(T_K^\star\) denote the first \(q_K\) items. A successful EAHR execution must return exactly this sequence. If fewer than \(K\) items have positive fusion scores, zero-score items are not appended.

During execution, retriever \(i\) exposes a stable exact prefix of its complete ranked list:

\[
P_i(d_i)=(\pi_i(1),\ldots,\pi_i(d_i)),
\qquad 0\le d_i\le n_i.
\]

Each ranked list is \texttt{Open}, \texttt{Exhausted}, or \texttt{Failed}. An \texttt{Open} list may extend its prefix from rank \(d_i+1\) but cannot revise earlier ranks; an \texttt{Exhausted} list is complete, with \(d_i=n_i\); a \texttt{Failed} list invalidates the EAHR request.

\subsection{Adaptive weighted-RRF fusion and ordered correctness}\label{adaptive-weighted-rrf-fusion-and-ordered-correctness}

Given the complete-list target and channel contract above, EAHR must decide how much of each channel ranking to obtain. For weighted RRF, this decision is based on the maximum contribution that unread ranks can still add. Fixed Top-\(L\) instead assigns these contributions zero, although they may still change Top-\(K\) membership or order.

Consider a checkpoint at which no ranked list has failed, and let \(\mathbf d=(d_i)_{i\in\mathcal C}\) be the current access depths. The items returned by retriever \(i\) and the union of all observed items are

\[
S_i(d_i)=\{\pi_i(r):1\le r\le d_i\},
\qquad
O(\mathbf d)=\bigcup_{i\in\mathcal C}S_i(d_i).
\]

An item not yet returned by an open list cannot occur before rank \(d_i+1\). Its remaining contribution from retriever \(i\) is therefore at most

\[
u_i(\mathbf d)=
\begin{cases}
g_i(d_i+1), & i\text{ is }\texttt{Open},\\
0, & i\text{ is }\texttt{Exhausted}.
\end{cases}
\]

For an observed item \(x\in O(\mathbf d)\), its known contributions give the lower bound

\[
L_{\mathbf d}(x)=
\sum_{i:x\in S_i(d_i)}g_i(r_i(x)).
\]

Open lists in which \(x\) has not appeared give the remaining upper bound

\[
U_{\mathbf d}(x)=L_{\mathbf d}(x)+
\sum_{\substack{i\text{ is }\texttt{Open}\\x\notin S_i(d_i)}}
u_i(\mathbf d).
\]

An unseen item may receive contributions from several open lists. Every item outside \(O(\mathbf d)\) therefore shares the upper bound

\[
B_{\mathbf d}=
\sum_{i\in\mathcal C}u_i(\mathbf d).
\]

Because every \(g_i\) is nonincreasing, for any \(x\in O(\mathbf d)\) and \(z\notin O(\mathbf d)\),

\[
L_{\mathbf d}(x)\le F(x)\le U_{\mathbf d}(x),
\qquad
F(z)\le B_{\mathbf d}.
\]

These bounds tighten monotonically as the ranked prefixes grow.

They determine the fused order one position at a time. Let \(A\) be the ordered prefix already fixed and \(R_{\mathbf d}=O(\mathbf d)\setminus A\). Choose \(x^\dagger\) from \(R_{\mathbf d}\) with the greatest lower bound, breaking ties by \(\prec_{\mathrm{id}}\). Define \((a,x)\succ(b,y)\) when \(a>b\), or when \(a=b\) and \(x\prec_{\mathrm{id}}y\). The item \(x^\dagger\) can be appended to \(A\) if

\[
\bigl(L_{\mathbf d}(x^\dagger),x^\dagger\bigr)
\succ
\bigl(U_{\mathbf d}(y),y\bigr),
\quad
\forall y\in R_{\mathbf d}\setminus\{x^\dagger\},
\]

and

\[
L_{\mathbf d}(x^\dagger)>B_{\mathbf d}.
\]

The first condition excludes every other observed item; the strict second excludes all unseen items, whose identifiers are not yet known. Thus, \(x^\dagger\) is the next item in the complete-list fusion order. Repeating the decision until \(q_K\) positions have been fixed yields

\[
T=T_K^\star.
\]

If the bounds do not yet determine the complete result, EAHR requests further ranks. When early determination is impossible, the contributing lists are exhausted and the same target is obtained from their complete fusion scores.

\subsection{Resumable exact ranked-list generation}\label{resumable-exact-ranked-list-generation}

Because EAHR does not know in advance how many ranks fusion will require, each retriever must be able to produce an exact prefix, pause, and later continue from the next rank without revising earlier output. PVS and PBM provide this behavior for dense vectors and compressed sparse postings.

\subsubsection{Dense ranked-list production with PVS}\label{dense-ranked-list-production-with-pvs}

Per-Vector Scalar Quantization (PVS) assigns each vector its own Int8 scale and uses the quantization error to bound its original float32 inner-product score. Items whose intervals determine their order require no raw-vector access; ambiguous items are evaluated exactly in float32.

Let the vector dimension be \(m\), and let \(v\in\mathbb R^m\) be an item vector. PVS stores an individual scale \(s_v>0\) and integer encoding

\[
z_v\in[-127,127]^m\cap\mathbb Z^m,
\]

with reconstruction \(\widehat v=s_vz_v\). For a nonzero vector, \(s_v\) is the positive float32 representation of \(\|v\|_\infty/127\). Each coordinate is encoded as \(\operatorname{round}(v_j/s_v)\) and clipped to \([-127,127]\). A zero vector uses \(s_v=1\) and an all-zero encoding. The query vector \(q\) is encoded in the same way as \(\widehat q=s_qz_q\), giving the quantized center score

\[
c(q,v)=s_qs_v\langle z_q,z_v\rangle.
\]

Let \(e_q=q-\widehat q\) and \(e_v=v-\widehat v\). Since

\[
q^\top v-\widehat q^\top\widehat v
=e_q^\top v+\widehat q^\top e_v,
\]

the Cauchy--Schwarz inequality gives

\[
\left|q^\top v-\widehat q^\top\widehat v\right|
\le
\|e_q\|_2\|v\|_2+
\|\widehat q\|_2\|e_v\|_2.
\]

Together with the finite-precision conditions in Appendix A.1, this residual bound gives a deterministic interval containing the authoritative float32 score. PVS scans the Int8 encodings to construct these intervals and processes candidates in decreasing order of interval upper bound. It reads a raw vector only when the remaining intervals do not determine the next rank, and retains exact scores and unprocessed candidates for continuation.

\subsubsection{Sparse ranked-list production with PBM}\label{sparse-ranked-list-production-with-pbm}

Posting Block-Max (PBM) partitions the item-identifier space into contiguous ranges and stores the maximum impact of each term within each range. Query weights turn these metadata into an upper bound on every sparse score in the range.

Let \(Q\) be the set of nonzero query terms, \(q_t\ge0\) the query weight of term \(t\), and \(p_t(x)\ge0\) the impact of term \(t\) on item \(x\). The sparse score is

\[
S(x)=\sum_{t\in Q}q_t p_t(x).
\]

For a contiguous range \(R\), define

\[
m_t(R)=\max_{x\in R}p_t(x),
\]

with \(m_t(R)=0\) when \(R\) contains no posting for term \(t\). PBM assigns the range the upper bound

\[
H(R)=\sum_{t\in Q}q_t m_t(R).
\]

Because all terms are nonnegative, every \(x\in R\) satisfies

\[
S(x)\le H(R).
\]

PBM expands ranges in decreasing order of \(H(R)\) and computes exact item scores from their postings. Once no unexpanded range can contain an item that precedes the current best scored item, PBM publishes the next rank. Unexpanded ranges and scored items are retained across batches. Appendix A.2 gives the finite-precision and block-max metadata conditions.

\subsection{Hierarchical merging and query execution}\label{hierarchical-merging-and-query-execution}

EAHR executes each query against a fixed set of Shard snapshots, whose filters and deletion states define the visible items. PVS and PBM produce tie-complete exact rankings within each Segment. For each retriever, these rankings are merged first within each Shard and then across Shards under the same score-and-identifier order, yielding one resumable request-level ranking. If the same item appears in multiple Segments, copies with consistent scores and versions are merged; conflicts invalidate the request.

The fusion session obtains the request-level dense and sparse rankings in batches and applies the bounds from Section 2.2 to the prefixes obtained so far. If the ordered Top-\(K\) is still unresolved, it requests further ranks and repeats the decision. Batch size and channel scheduling may change the work performed, but not the result, provided every layer maintains the exact-prefix contract in Section 2.1.

EAHR keeps the standard retrieval interface. The caller supplies a query, \(K\), and fusion parameters and receives at most \(K\) ranked items. The caller does not set a per-retriever Top-\(L\) cutoff or receive intermediate ranking state.

\section{Experimental Setup}\label{experimental-setup}

\subsection{Research questions and common configuration}\label{research-questions-and-common-configuration}

The evaluation separates the retrieval target from the work needed to obtain it. RQ1 asks whether weighted RRF over complete ranked lists is a competitive target and whether a fixed Top-\(L\) chosen from historical queries transfers to new queries and later corpus snapshots. RQ2 verifies that EAHR returns the complete-list ordered Top-\(K\) and examines how queries and corpus snapshots alter the channel rankings and, in turn, the required access depth. RQ3 compares EAHR with exhaustive execution in work and latency at both aggregate and individual-query levels. RQ4 measures the contribution of PVS and PBM by replacing each with alternative exact rank generators.

All real-query experiments use one dense and one sparse retriever. Weighted RRF uses rank constant \(k=60\), equal retriever weights, and a fixed item-identifier order for ties.

\subsection{Test collections, temporal snapshots, and retrieval representations}\label{test-collections-temporal-snapshots-and-retrieval-representations}

The static evaluation uses the BEIR test splits of NFCorpus (3,633 items, 323 queries), SciFact (5,183, 300), and TREC-COVID (171,332, 50) \citep{boteva2016nfcorpus, wadden2020scifact, roberts2021treccovid, thakur2021beir}. TREC-DL 2019 and 2020 use 43 and 54 judged queries over the same 8,841,823-passage MS MARCO corpus and index \citep{bajaj2016msmarco, craswell2020trecdl2019, craswell2021trecdl2020}. All five evaluations use the complete corpus and all official test queries.

Temporal transfer is evaluated on a separate series of five CORD-19 snapshots released between April 10 and July 16, 2020 \citep{roberts2021treccovid, nist_trec_covid_round5}. After deduplication by CORD UID, the corpus grows from 51,045 items in Round 1 to 191,175 in Round 5. The same 30 topics are evaluated on every snapshot with Chronological qrels \citep{nist_trec_covid_data}.

Dense retrieval uses BAAI bge-small-en-v1.5 to encode queries and items as 384-dimensional, L2-normalized float32 vectors with a 512-token input limit \citep{xiao2024cpack}. Sparse retrieval uses the nonnegative bm25-impact-v1 representation. We generate it with Pyserini 2.3.0/Anserini on JDK 21.0.10 \citep{lin2021pyserini}. Its English analyzer applies Porter stemming and stopword removal, with BM25 parameters \(k_1=0.9\) and \(b=0.4\) \citep{robertson2009bm25}. The dense encoder remains fixed across snapshots, while sparse corpus statistics are recomputed for each snapshot.

\subsection{Retrieval effectiveness and fixed-depth transfer}\label{retrieval-effectiveness-and-fixed-depth-transfer}

The static comparison covers dense-only retrieval, positive-score sparse-only retrieval, fixed Top-\(L\) weighted RRF, and complete-list weighted RRF. Fixed Top-\(L\) fusion uses

\[
L\in\{10,20,50,100,200,500,1000,2000,5000\},
\]

whereas complete-list fusion exhausts both rankings before applying the same rule. Because EAHR reproduces the complete-list result, it has no separate effectiveness row. All methods return up to 100 items; complete-list fusion additionally retains rank 101 to verify the Top-100 boundary.

Retrieval effectiveness is computed with \texttt{ir\_measures\ 0.4.3} using nDCG@10, MRR@10, and Recall@100. When \(L<100\), the fixed-window candidate union may contain fewer than 100 items; these Recall@100 values are therefore capacity-limited and are not used to infer fusion-quality loss. Static differences from complete-list fusion use 95\% confidence intervals from 10,000 paired query-level bootstrap replicates. Candidate coverage, membership agreement, and ordered agreement use Wilson 95\% confidence intervals.

The temporal experiment tests whether a depth selected on Round 1 transfers to held-out topics and later corpus snapshots. Topic \(q\) is assigned to fold \((q-1)\bmod 5\). For each fold, the other 24 Round 1 topics select the \(L\) with the highest mean nDCG@10, with ties resolved toward the smaller depth. This value is then fixed for the six held-out topics across Rounds 1--5. Round 1 tests transfer to unseen topics on the same corpus; Rounds 2--5 additionally test transfer after corpus updates.

As a descriptive diagnostic, we also report the best tested \(L\) for each static collection and temporal round. This is the grid depth with the highest test-set mean nDCG@10, with ties again resolved toward the smaller value; it is not treated as a deployable selection rule.

For each held-out topic, we subtract the complete-list score from the score obtained at its fold-selected \(L\). The 95\% confidence intervals use 10,000 nested bootstrap replicates that resample the Round 1 selection topics, reselect \(L\), resample the held-out topics, and carry the new selection across all five snapshots.

\subsection{EAHR exactness and adaptive access}\label{eahr-exactness-and-adaptive-access}

We evaluate EAHR with \(K=20\) on topics 1--30 in each of the five CORD-19 snapshots, yielding 150 query--snapshot runs. Every ordered result is compared position by position with complete-list fusion. For each retriever \(i\), we record access depth \(d_i\), complete-list length \(n_i\), read ratio \(d_i/n_i\), and exhaustion, together with membership, order, and tie-order mismatches.

To determine whether read ratios vary mainly by query or corpus snapshot, we apply a two-factor sum-of-squares decomposition to \(\log_{10}(d_i/n_i)\), separately for dense and sparse retrieval. The decomposition attributes the variation to query, snapshot, and query--snapshot interaction components.

Controlled rank streams isolate the effects of collection size and cross-retriever rank relationship. We use collection sizes of \(100\mathrm{K}\), \(1\mathrm{M}\), and \(5\mathrm{M}\), with seeds 1729, 2027, and 65537. Except in the all-tied setting, the first list is a random permutation. The second list is produced by permuting within contiguous 32-item blocks for highly correlated rankings; placing 500 items from the first list's Top-1000 and 500 outside items in its own Top-1000 for partial overlap; drawing an independent permutation; or reversing the first list for anti-correlation. In the all-tied setting, every item receives the same score in both lists. The three sizes, five relationships, and three seeds produce 45 instances.

All controlled instances use \(K=20\). Both lists advance to the same depth, and certification is checked every 16 ranks. We report the first checkpoint at which the ordered Top-20 is fixed, the corresponding read ratio, and whether either list is exhausted. These synthetic rankings evaluate fusion-level access behavior, not retrieval effectiveness or system latency.

\subsection{Execution cost and rank-generator comparisons}\label{execution-cost-and-rank-generator-comparisons}

The end-to-end experiment compares EAHR with two exhaustive baselines, all returning the same ordered Top-20. Same-producer exhaustive execution uses the same PVS, PBM, scheduler, and batched continuation as EAHR but runs both rankings to exhaustion. Its comparison with EAHR measures the work avoided by adaptive stopping and covers NFCorpus, SciFact, and TREC-COVID. Exhaustive batch execution generates both complete rankings through an optimized parallel path and is compared with EAHR on all five static collections.

We measure ranked access depth and ratio, dense float32 score evaluations, sparse posting visits, scored sparse items, and end-to-end latency at p50, p95, and p99. Failed executions are reported separately and excluded from cost summaries; all successful executions remain in the latency distribution.

In the frozen implementation, each scheduling step reads at most 16 ranks from one retriever. The scheduler selects the retriever with the larger maximum weighted-RRF contribution at its next unread rank, while a fairness rule advances a retriever after 16 skipped steps. For Top-20, certification is checked after every batch through the first 80 returned ranks. Later checks occur after one-eighth of cumulative ranked access, with the interval limited to 64--1024 ranks; exhaustion of either list triggers an immediate check.

EAHR and each exhaustive baseline are measured in warm-cache pairs. Each query receives two warm-up runs and five measured runs per plan, with execution order rotated across repetitions. Exhaustive/EAHR latency ratios are geometrically averaged first within each query and then across queries. Confidence intervals use 10,000 query-level bootstrap replicates with seed 20260730.

Rank-generator comparisons cover NFCorpus, SciFact, TREC-COVID, and TREC-DL 2020. Using PVS+PBM as the reference, each experiment replaces one generator: PVS is compared with scalar-quantization intervals and exhaustive float32 scanning, and PBM with complete exact sparse ranking generation. The four configurations are measured in three Williams-balanced rounds, with two consecutive observations per configuration in each round and separate Qdrant processes for query shards. Each alternative is paired with PVS+PBM by query; 95\% confidence intervals use 10,000 query-level bootstrap replicates with seed 20260730.

\subsection{Execution environment}\label{execution-environment}

Latency experiments run as local processes on a 14-core Apple M4 Pro (arm64) with 24 GB of memory and macOS 26.5.2. The system uses Qdrant \texttt{v1.18.2}, built in \texttt{-\/-release} mode with \texttt{rustc\ 1.99.0-nightly}. TREC-DL experiments use one Shard and eight immutable indexed Segments. Timing starts from precomputed query representations and excludes query encoding and downstream RAG processing.

\section{Results}\label{results}

\subsection{Retrieval effectiveness and fixed-depth effects}\label{retrieval-effectiveness-and-fixed-depth-effects}

EAHR uses the ordered result of complete-list weighted RRF as its retrieval target. Complete-list fusion outperformed both dense-only and sparse-only retrieval on NFCorpus, SciFact, and TREC-COVID, and lay between them on TREC-DL 2019 and 2020. The largest gain appeared on TREC-COVID, where nDCG@10 increased from 0.6649 for dense-only retrieval to 0.8059. No method was best on all five collections.

\begin{figure}[!ht]
\centering
\includegraphics[width=\columnwidth]{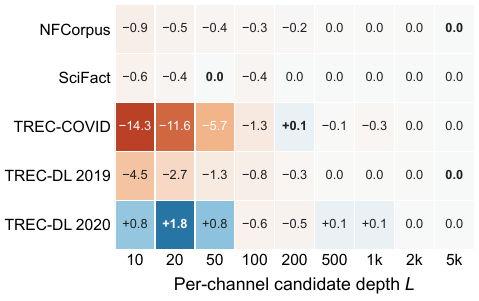}
\caption{Fixed Top-\(L\) nDCG@10 relative to complete-list weighted RRF. Cells show mean per-query differences in points; positive values favor fixed Top-\(L\), and bold marks the best tested \(L\) for each collection.}
\Description{A five-by-nine heatmap of nDCG at 10 differences between fixed Top-L and complete-list weighted RRF. Rows are test collections and columns are candidate depths. Each cell contains its signed numeric difference; bold values identify the best tested depth for each collection.}
\label{fig:fixed-depth-effectiveness}
\end{figure}

Figure \ref{fig:fixed-depth-effectiveness} compares fixed Top-\(L\) fusion with the same complete-list target. On TREC-COVID, nDCG@10 increased from 0.6624 at \(L=10\) to 0.7925 at \(L=100\), while complete-list fusion reached 0.8059. TREC-DL 2020 behaved differently: nDCG@10 peaked at 0.6424 at \(L=20\), fell to 0.6179 at \(L=100\), and remained near 0.624 at larger depths. Increasing the candidate depth therefore did not consistently improve retrieval effectiveness. Across the tested grid, the highest mean nDCG@10 occurred at \(L=5000\), 50, 200, 5000, and 20 for NFCorpus, SciFact, TREC-COVID, TREC-DL 2019, and TREC-DL 2020, respectively. TREC-DL 2019 and 2020 use the same corpus, representations, indexes, and fusion settings but different query sets; their preferred depths were \(L=5000\) and \(L=20\). Thus, even under the same corpus and system configuration, the depth favored by fixed-window fusion can change with the queries.

Similar nDCG@10 values did not imply identical rankings. At \(L=100\), the two truncated lists together contained every item in the complete-list Top-20 on all five collections, but exact ordered agreement was only 44.9\%, 41.7\%, 16.0\%, 44.2\%, and 31.5\%, respectively. An item observed in one list may still appear beyond the cutoff in the other, where its omitted contribution can change the fusion score and position. Fixed \(L\) therefore couples execution cost to both retrieval effectiveness and the returned ranking.

\subsection{Adaptive access across queries and corpus snapshots}\label{adaptive-access-across-queries-and-corpus-snapshots}

\begin{figure}[!ht]
\centering
\includegraphics[width=\columnwidth]{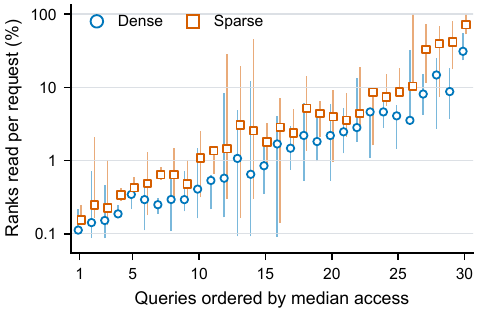}
\caption{EAHR ranked-list access for 30 queries across five CORD-19 snapshots. Queries are ordered by median access. Dense and sparse markers show the median percentage of each complete ranking read; vertical lines show the minimum and maximum percentages for the same query across the five snapshots.}
\Description{For each of 30 queries, dense and sparse markers show the median fraction of each ranked list read by EAHR, while vertical ranges show the minimum and maximum over five corpus snapshots.}
\label{fig:eahr-request-adaptation}
\end{figure}

EAHR matched complete-list fusion at every Top-20 position in all 150 query and snapshot combinations. The access needed to determine these results varied with both the query and the corpus snapshot. Figure \ref{fig:eahr-request-adaptation} shows both forms of variation. Across the 30 queries, median dense and sparse read ratios differed by more than two orders of magnitude. Holding the query fixed, all 30 queries changed their dense and sparse access depths across the five snapshots. From Round 1 to Round 5, 23 queries required greater absolute depth and seven required less depth in both channels.

The variance decomposition separates these two sources. Query identity accounted for 75.2\% of the dense variation and 76.9\% of the sparse variation in log read ratios. Snapshot main effects accounted for 1.3\% and 2.5\%, while query and snapshot interactions accounted for 23.4\% and 20.6\%. Together, the snapshot and interaction components accounted for 24.8\% of dense variation and 23.1\% of sparse variation. Query differences therefore dominated, but corpus updates still changed the access required by individual queries rather than producing a uniform shift across all requests.

The controlled experiment explains why queries can require such different depths. Correlated rankings fixed the Top-20 at depth 32 for all three collection sizes, whereas partial-overlap rankings stopped at a median depth of 1264. Independent rankings required 64.5\% to 77.9\% of each list, and all nine anti-correlated instances exhausted both lists. When the rankings agree near the top, leading items receive contributions from both retrievers early. As the rankings diverge, competing items remain deeper in the two lists, delaying determination of the fused order.

The temporal effectiveness results show why a fixed depth is difficult to transfer. Using the depths selected from Round 1, mean nDCG@10 was below complete-list fusion in all five rounds.

In Round 5, the difference was \(-0.02721\), with a 95\% confidence interval of \([-0.07031,-0.00561]\). If each round was instead allowed to choose its best tested depth after evaluation, the selected value changed from 50 to 100, 200, 200, and 5000. A depth chosen from earlier queries and an earlier corpus snapshot may therefore become unsuitable as either changes.

\subsection{Execution efficiency and query-level tail}\label{execution-efficiency-and-query-level-tail}

\begin{figure}[!ht]
\centering
\includegraphics[width=\columnwidth]{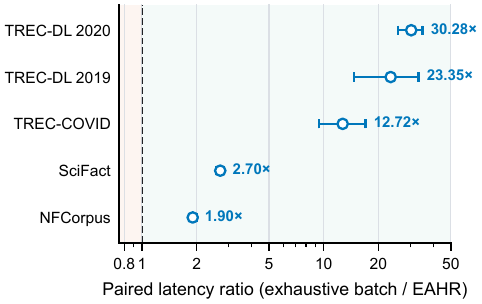}
\caption{Exhaustive-batch/EAHR latency ratios paired by query. Points show geometric means and bars show 95\% confidence intervals; values above one favor EAHR.}
\Description{Paired exhaustive-batch-to-EAHR latency ratios across five test collections, with geometric means and 95 percent confidence intervals.}
\label{fig:eahr-aggregate-efficiency}
\end{figure}

Figure \ref{fig:eahr-aggregate-efficiency} compares EAHR with exhaustive batch execution, which generates both complete rankings through an optimized parallel path. We also compare EAHR with same-producer exhaustive execution. This baseline uses the same PVS, PBM, and batched continuation as EAHR but exhausts both rankings.

EAHR was faster in all five exhaustive-batch comparisons. The paired geometric-mean ratios ranged from 1.90 on NFCorpus to 30.28 on TREC-DL 2020, with every confidence interval above one. Same-producer exhaustive execution was 2.36 to 16.42 times slower. Because this baseline changes only the stopping behavior, the difference measures the work avoided after the ordered Top-20 was fixed.

The logical access and physical counters show where these savings arose. On NFCorpus, EAHR exhausted the sparse ranking but read a median 5.73\% of the dense ranking. On TREC-COVID, it read 1.18\% and 1.95\% of the dense and sparse rankings. Median dense float32 evaluations on NFCorpus, SciFact, and TREC-COVID fell from 3633, 5183, and 171332 under exhaustive execution to 896, 864, and 10152 under EAHR. PBM posting visits and scored nonzero items were unchanged, so early stopping reduced the sparse ranks delivered to fusion rather than the posting work needed to produce them.

On TREC-DL 2019 and 2020, median latency was 0.929 and 0.923 s for EAHR, compared with 19.788 and 19.404 s for exhaustive batch execution. Execution order affected the size of the difference: the exhaustive/EAHR ratios were 57.10 and 78.19 when EAHR ran first, and 9.48 and 11.75 when exhaustive batch ran first. Both execution-order strata favored EAHR.

The aggregate result did not hold for every query. EAHR's p95 and p99 latencies were 16.665 and 504.214 s on TREC-DL 2019, and 6.748 and 14.219 s on TREC-DL 2020. For TREC-DL 2019 queries 855410 and 443396, EAHR's median latency was respectively 25.603 and 4.694 times that of exhaustive batch execution. Both still returned the correct ordered Top-20, but deep ranked access and repeated continuation made the incremental path more expensive. The latency benefit is therefore query dependent.

\FloatBarrier

\subsection{Rank-generator ablation}\label{rank-generator-ablation}

To measure the contribution of the rank generators themselves, we kept EAHR's fusion target and stopping process unchanged and replaced one generator at a time. Table~\ref{tab:4-1} reports the paired geometric-mean latency ratio of each alternative to PVS or PBM. Ratios above one favor PVS or PBM. None of the 95\% confidence intervals crossed one.

\begin{table}[H]
\centering
\footnotesize
\setlength{\tabcolsep}{3pt}
\caption{Paired geometric-mean alternative/reference latency ratios.}
\label{tab:4-1}
\begin{tabular}[]{@{}lrrr@{}}
\toprule\noalign{}
Collection & Scalar/PVS & Scan/PVS & Sparse/PBM \\
\midrule\noalign{}
NFCorpus & 1.327 & 0.915 & 1.027 \\
SciFact & 1.302 & 0.963 & 1.126 \\
TREC-COVID & 1.757 & 1.117 & 2.747 \\
TREC-DL 2020 & 2.388 & 52.819 & 2.108 \\
\bottomrule\noalign{}
\end{tabular}
\end{table}

Scalar-quantization intervals were 1.302--2.388 times slower than PVS. Full float32 scanning was faster on NFCorpus and SciFact, but PVS became faster on TREC-COVID and reached a 52.819-fold advantage on TREC-DL 2020. PVS reduced median float32 evaluations by 59.3\%--86.4\% relative to scalar-quantization intervals and by 75.3\%--99.97\% relative to full scanning. On TREC-DL 2020, the median fell from 8,841,823 evaluations to 2,664. On the two smaller collections, the saved evaluations did not offset interval construction, quantized scanning, and queue maintenance.

Complete sparse ranking was 1.027--2.747 times slower than PBM. The two methods scored the same number of nonzero items. PBM's advantage came from producing the ranking incrementally in range-bound order rather than constructing it in full before fusion.

\FloatBarrier

\section{Discussion}\label{discussion}

\subsection{From fixed retrieval depth to exact adaptive execution}\label{from-fixed-retrieval-depth-to-exact-adaptive-execution}

Fixed Top-\(L\) is often introduced as an execution budget, but it also determines which rank contributions enter fusion. Changing \(L\) can therefore change the returned items and their order, not only the work performed. Truncated fusion is a well-defined retrieval function when chosen deliberately. Our concern is the use of \(L\) as an execution approximation when complete-list fusion is the intended target. Our experiments found no fixed depth that served reliably across collections, queries, and corpus snapshots. The depth with the highest effectiveness changed across these settings, and even large windows often failed to reproduce the complete-list ordered Top-20.

EAHR replaces this coupling with a fixed result contract. For a given query, corpus snapshot, and fusion configuration, complete-list weighted RRF defines the ordered Top-\(K\), while channel depth remains internal execution state. A new query or snapshot may change both the target ranking and the evidence needed to determine it, but the rule defining the correct response remains unchanged. The system no longer selects \(L\) to trade result quality against execution cost. It reduces work subject to preserving the declared complete-list result.

This contract changes the execution of the whole hybrid-retrieval request. At query start, the system does not already possess two complete lists ordered for the current query. Dense and sparse retrieval must produce further exact ranks as needed, preserve their state across batches, and merge their outputs under one snapshot. EAHR coordinates these operations: PVS and PBM generate resumable exact rankings, hierarchical merging forms request-level rankings, and weighted RRF determines when the ordered Top-\(K\) is fixed.

\subsection{Logical ranking evidence and physical execution cost}\label{logical-ranking-evidence-and-physical-execution-cost}

EAHR adapts the number of exact ranks delivered to fusion, but rank delivery is not equivalent to the physical work performed. Before publishing a ranked prefix, a retriever may scan quantized records, decode postings, score candidates, and maintain state for later continuation. Fusion adds bound updates, merging, scheduling, and repeated continuation. The two exhaustive baselines in Section 4.3 expose different parts of this cost. Same-producer exhaustive execution measures the work avoided after the ordered Top-\(K\) is fixed, whereas exhaustive batch execution compares EAHR with a separately optimized complete-ranking plan. Access depth therefore measures the ranking evidence required by fusion, not the total cost of obtaining it.

The dense and sparse results illustrate this distinction in different ways. Early stopping substantially reduced dense float32 evaluations, but PVS still scanned Int8 encodings, constructed score intervals, and maintained ranking state. Full float32 scanning was therefore faster on NFCorpus and SciFact, while PVS became faster as the collection grew and avoided exact scoring outweighed its overhead. Sparse retrieval showed a different boundary. PBM was faster than constructing a complete sparse ranking on all four evaluated collections, yet early fusion stopping did not reduce posting visits or scored nonzero items once PBM was already in use. A short exact prefix may still require processing much of the relevant posting data. These results agree with prior dense and sparse retrieval studies showing that pruning benefits depend on the work required to maintain and evaluate bounds \citep{abuzaid2019exactmips, carlson2025superblock, carlson2026superblock}.

The relationship between the channel rankings determines how much ranking evidence is needed in the first place. When dense and sparse rankings agree near the top, shared items accumulate enough contribution for early determination. Anti-correlated rankings leave plausible competitors deep in opposite lists and may force both lists to be exhausted. Queries 855410 and 443396 exhibited this difficult behavior: deep access and repeated continuation made EAHR slower than exhaustive batch execution, although both returned the same ordered Top-20. For every successful execution, the result contract remains unchanged; the performance benefit depends jointly on the evidence required by the query and the physical cost of producing it.

\subsection{Scope and deployment implications}\label{scope-and-deployment-implications}

The EAHR result contract and the reported performance have different scopes. The ordered-correctness argument applies to successful executions in which each channel exposes a resumable exact ranking under one fixed snapshot, rank contributions are nonnegative and nonincreasing, and deterministic ties are preserved. Our experiments instantiate this contract with one dense retriever, one sparse retriever, equal weights, and \(K=20\). The latency evidence is narrower: it comes from warm-cache, sequential local execution on an Apple M4 Pro, with the scale experiments using one Shard and eight Segments. These measurements characterize the evaluated realization; they do not establish universal speedups, cold-cache or concurrent performance, index and storage costs, or distributed correctness.

For a RAG caller, EAHR changes where the execution policy resides. A fixed Top-\(L\) interface asks the caller or system configuration to decide in advance how deeply each retriever contributes to fusion. With EAHR, the fusion configuration is fixed and the request specifies the query and \(K\); the retrieval layer determines the channel depths internally and returns the same ordered Top-\(K\) interface. Reranking, context construction, and generation do not need access to channel prefixes, continuation state, or stopping decisions. EAHR therefore preserves the standard Top-\(K\) interface while moving channel-depth decisions inside the retrieval system.

Extending this design across nodes requires the result contract to become a distributed consistency contract. The current implementation can merge multiple local Shards, but remote replicas, cross-node Shards, and failover must preserve a common visible snapshot, item versions, global tie order, prefix continuity, and failure semantics. The same principle applies to other extensions. More retrievers and other fusion rules can be supported when their rank contributions remain nonnegative and nonincreasing, while a cost-aware scheduler could choose among result-equivalent exact rank generators as a query progresses. In each case, the execution plan may change, but every successful response must still match the declared complete-list fusion ranking in both items and order.

\section{Conclusion}\label{conclusion}

Fixed Top-\(L\) gives a single parameter two roles in hybrid retrieval: it limits
how far each retriever runs and determines which rank contributions enter
fusion. When the intended target is weighted RRF over complete rankings, a
fixed prefix cannot certify that the ordered Top-\(K\) is final, and a depth
selected from past queries may not transfer as the query or corpus changes.
EAHR separates these roles by fixing the complete-list ordered Top-\(K\) as the
result contract and treating channel depth as internal execution state. In our
realization, PVS and PBM produce resumable exact dense and sparse rankings,
while fusion bounds contributions from unread ranks and requests more only
while they can change Top-\(K\) membership or order. A successful request
therefore either stops with the same ordered result as complete-list fusion or
continues safely to list exhaustion, without requiring a preset Top-\(L\).

Complete-list weighted RRF remained competitive across five test collections,
while the best tested fixed depth varied across collections, query sets, and
corpus snapshots. EAHR matched the complete-list ordered Top-20 in all 150
query--snapshot combinations. Under the reported warm-cache, interleaved,
order-balanced protocol, the paired geometric-mean exhaustive/EAHR latency
ratios reached 23.35 on TREC-DL 2019 and 30.28 on TREC-DL 2020. The benefit was
not universal: all nine anti-correlated instances exhausted both rankings, and
queries 855410 and 443396 were slower with EAHR because they required deep
access and repeated continuation. EAHR does not guarantee a speedup for every
request. It fixes the retrieval result without a preset Top-\(L\) and allows
execution work to adapt when the current rankings provide sufficient evidence.

\bibliographystyle{ACM-Reference-Format}
\bibliography{references}


\begin{thebibliography}{30}


\ifx \showCODEN    \undefined \def \showCODEN     #1{\unskip}     \fi
\ifx \showISBNx    \undefined \def \showISBNx     #1{\unskip}     \fi
\ifx \showISBNxiii \undefined \def \showISBNxiii  #1{\unskip}     \fi
\ifx \showISSN     \undefined \def \showISSN      #1{\unskip}     \fi
\ifx \showLCCN     \undefined \def \showLCCN      #1{\unskip}     \fi
\ifx \shownote     \undefined \def \shownote      #1{#1}          \fi
\ifx \showarticletitle \undefined \def \showarticletitle #1{#1}   \fi
\ifx \showURL      \undefined \def \showURL       {\relax}        \fi
\providecommand\bibfield[2]{#2}
\providecommand\bibinfo[2]{#2}
\providecommand\natexlab[1]{#1}
\providecommand\showeprint[2][]{arXiv:#2}

\bibitem[Abuzaid et~al\mbox{.}(2019)]%
        {abuzaid2019exactmips}
\bibfield{author}{\bibinfo{person}{Firas Abuzaid}, \bibinfo{person}{Geet Sethi}, \bibinfo{person}{Peter Bailis}, {and} \bibinfo{person}{Matei Zaharia}.} \bibinfo{year}{2019}\natexlab{}.
\newblock \showarticletitle{To Index or Not to Index: Optimizing Exact Maximum Inner Product Search}. In \bibinfo{booktitle}{\emph{2019 IEEE 35th International Conference on Data Engineering}}. \bibinfo{pages}{1250--1261}.
\newblock
\href{https://doi.org/10.1109/ICDE.2019.00114}{doi:\nolinkurl{10.1109/ICDE.2019.00114}}


\bibitem[Bajaj et~al\mbox{.}(2016)]%
        {bajaj2016msmarco}
\bibfield{author}{\bibinfo{person}{Payal Bajaj}, \bibinfo{person}{Daniel Campos}, \bibinfo{person}{Nick Craswell}, \bibinfo{person}{Li Deng}, \bibinfo{person}{Jianfeng Gao}, \bibinfo{person}{Xiaodong Liu}, \bibinfo{person}{Rangan Majumder}, \bibinfo{person}{Andrew McNamara}, \bibinfo{person}{Bhaskar Mitra}, \bibinfo{person}{Tri Nguyen}, \bibinfo{person}{Mir Rosenberg}, \bibinfo{person}{Xia Song}, \bibinfo{person}{Alina Stoica}, \bibinfo{person}{Saurabh Tiwary}, {and} \bibinfo{person}{Tong Wang}.} \bibinfo{year}{2016}\natexlab{}.
\newblock \showarticletitle{{MS MARCO}: A Human Generated Machine Reading Comprehension Dataset}.
\newblock \bibinfo{journal}{\emph{arXiv preprint arXiv:1611.09268}} (\bibinfo{year}{2016}).
\newblock
\showeprint[arxiv]{1611.09268}


\bibitem[Boteva et~al\mbox{.}(2016)]%
        {boteva2016nfcorpus}
\bibfield{author}{\bibinfo{person}{Vera Boteva}, \bibinfo{person}{Demian Gholipour}, \bibinfo{person}{Artem Sokolov}, {and} \bibinfo{person}{Stefan Riezler}.} \bibinfo{year}{2016}\natexlab{}.
\newblock \showarticletitle{A Full-Text Learning to Rank Dataset for Medical Information Retrieval}. In \bibinfo{booktitle}{\emph{Advances in Information Retrieval}}. \bibinfo{publisher}{Springer}, \bibinfo{pages}{716--722}.
\newblock
\href{https://doi.org/10.1007/978-3-319-30671-1_58}{doi:\nolinkurl{10.1007/978-3-319-30671-1_58}}


\bibitem[Broder et~al\mbox{.}(2003)]%
        {broder2003wand}
\bibfield{author}{\bibinfo{person}{Andrei~Z. Broder}, \bibinfo{person}{David Carmel}, \bibinfo{person}{Michael Herscovici}, \bibinfo{person}{Aya Soffer}, {and} \bibinfo{person}{Jason Zien}.} \bibinfo{year}{2003}\natexlab{}.
\newblock \showarticletitle{Efficient Query Evaluation Using a Two-Level Retrieval Process}. In \bibinfo{booktitle}{\emph{Proceedings of the Twelfth International Conference on Information and Knowledge Management}}. \bibinfo{pages}{426--434}.
\newblock
\href{https://doi.org/10.1145/956863.956944}{doi:\nolinkurl{10.1145/956863.956944}}


\bibitem[Bruch et~al\mbox{.}(2024a)]%
        {bruch2024fusion}
\bibfield{author}{\bibinfo{person}{Sebastian Bruch}, \bibinfo{person}{Siyu Gai}, {and} \bibinfo{person}{Amir Ingber}.} \bibinfo{year}{2024}\natexlab{a}.
\newblock \showarticletitle{An Analysis of Fusion Functions for Hybrid Retrieval}.
\newblock \bibinfo{journal}{\emph{ACM Transactions on Information Systems}} \bibinfo{volume}{42}, \bibinfo{number}{1}, Article \bibinfo{articleno}{20} (\bibinfo{year}{2024}), \bibinfo{numpages}{35}~pages.
\newblock
\href{https://doi.org/10.1145/3596512}{doi:\nolinkurl{10.1145/3596512}}


\bibitem[Bruch et~al\mbox{.}(2024b)]%
        {bruch2024bridging}
\bibfield{author}{\bibinfo{person}{Sebastian Bruch}, \bibinfo{person}{Franco~Maria Nardini}, \bibinfo{person}{Amir Ingber}, {and} \bibinfo{person}{Edo Liberty}.} \bibinfo{year}{2024}\natexlab{b}.
\newblock \showarticletitle{Bridging Dense and Sparse Maximum Inner Product Search}.
\newblock \bibinfo{journal}{\emph{ACM Transactions on Information Systems}} \bibinfo{volume}{42}, \bibinfo{number}{6}, Article \bibinfo{articleno}{151} (\bibinfo{year}{2024}), \bibinfo{numpages}{38}~pages.
\newblock
\href{https://doi.org/10.1145/3665324}{doi:\nolinkurl{10.1145/3665324}}


\bibitem[Carlson et~al\mbox{.}(2025)]%
        {carlson2025superblock}
\bibfield{author}{\bibinfo{person}{Parker Carlson}, \bibinfo{person}{Wentai Xie}, \bibinfo{person}{Shanxiu He}, {and} \bibinfo{person}{Tao Yang}.} \bibinfo{year}{2025}\natexlab{}.
\newblock \showarticletitle{Dynamic Superblock Pruning for Fast Learned Sparse Retrieval}. In \bibinfo{booktitle}{\emph{Proceedings of the 48th International ACM SIGIR Conference on Research and Development in Information Retrieval}}. \bibinfo{pages}{3004--3009}.
\newblock
\showeprint[arxiv]{2504.17045}
\href{https://doi.org/10.1145/3726302.3730183}{doi:\nolinkurl{10.1145/3726302.3730183}}


\bibitem[Carlson et~al\mbox{.}(2026)]%
        {carlson2026superblock}
\bibfield{author}{\bibinfo{person}{Parker Carlson}, \bibinfo{person}{Wentai Xie}, \bibinfo{person}{Rohil Shah}, {and} \bibinfo{person}{Tao Yang}.} \bibinfo{year}{2026}\natexlab{}.
\newblock \showarticletitle{Efficient Sparse Retrieval with Lightweight Superblock Pruning}. In \bibinfo{booktitle}{\emph{Proceedings of the 49th International ACM SIGIR Conference on Research and Development in Information Retrieval}}. \bibinfo{pages}{133--144}.
\newblock
\showeprint[arxiv]{2602.02883}
\href{https://doi.org/10.1145/3805712.3809739}{doi:\nolinkurl{10.1145/3805712.3809739}}


\bibitem[Cormack et~al\mbox{.}(2009)]%
        {cormack2009rrf}
\bibfield{author}{\bibinfo{person}{Gordon~V. Cormack}, \bibinfo{person}{Charles L.~A. Clarke}, {and} \bibinfo{person}{Stefan Buettcher}.} \bibinfo{year}{2009}\natexlab{}.
\newblock \showarticletitle{Reciprocal Rank Fusion Outperforms Condorcet and Individual Rank Learning Methods}. In \bibinfo{booktitle}{\emph{Proceedings of the 32nd International ACM SIGIR Conference on Research and Development in Information Retrieval}}. \bibinfo{pages}{758--759}.
\newblock
\href{https://doi.org/10.1145/1571941.1572114}{doi:\nolinkurl{10.1145/1571941.1572114}}


\bibitem[Craswell et~al\mbox{.}(2021)]%
        {craswell2021trecdl2020}
\bibfield{author}{\bibinfo{person}{Nick Craswell}, \bibinfo{person}{Bhaskar Mitra}, \bibinfo{person}{Emine Yilmaz}, {and} \bibinfo{person}{Daniel Campos}.} \bibinfo{year}{2021}\natexlab{}.
\newblock \showarticletitle{Overview of the {TREC} 2020 Deep Learning Track}.
\newblock \bibinfo{journal}{\emph{arXiv preprint arXiv:2102.07662}} (\bibinfo{year}{2021}).
\newblock
\showeprint[arxiv]{2102.07662}


\bibitem[Craswell et~al\mbox{.}(2020)]%
        {craswell2020trecdl2019}
\bibfield{author}{\bibinfo{person}{Nick Craswell}, \bibinfo{person}{Bhaskar Mitra}, \bibinfo{person}{Emine Yilmaz}, \bibinfo{person}{Daniel Campos}, {and} \bibinfo{person}{Ellen~M. Voorhees}.} \bibinfo{year}{2020}\natexlab{}.
\newblock \showarticletitle{Overview of the {TREC} 2019 Deep Learning Track}.
\newblock \bibinfo{journal}{\emph{arXiv preprint arXiv:2003.07820}} (\bibinfo{year}{2020}).
\newblock
\showeprint[arxiv]{2003.07820}


\bibitem[Ding and Suel(2011)]%
        {ding2011bmw}
\bibfield{author}{\bibinfo{person}{Shuai Ding} {and} \bibinfo{person}{Torsten Suel}.} \bibinfo{year}{2011}\natexlab{}.
\newblock \showarticletitle{Faster Top-k Document Retrieval Using Block-Max Indexes}. In \bibinfo{booktitle}{\emph{Proceedings of the 34th International ACM SIGIR Conference on Research and Development in Information Retrieval}}. \bibinfo{pages}{993--1002}.
\newblock
\href{https://doi.org/10.1145/2009916.2010048}{doi:\nolinkurl{10.1145/2009916.2010048}}


\bibitem[{Elastic}({[n.\,d.]})]%
        {elastic_rrf_docs}
\bibfield{author}{\bibinfo{person}{{Elastic}}.} \bibinfo{year}{[n.\,d.]}\natexlab{}.
\newblock \bibinfo{title}{Reciprocal Rank Fusion}.
\newblock \bibinfo{howpublished}{\url{https://www.elastic.co/docs/reference/elasticsearch/rest-apis/reciprocal-rank-fusion}}.
\newblock
\shownote{Accessed 2026-07-29}.
\newblock


\bibitem[Fagin et~al\mbox{.}(2003a)]%
        {fagin2003medrank}
\bibfield{author}{\bibinfo{person}{Ronald Fagin}, \bibinfo{person}{Ravi Kumar}, {and} \bibinfo{person}{D. Sivakumar}.} \bibinfo{year}{2003}\natexlab{a}.
\newblock \showarticletitle{Efficient Similarity Search and Classification via Rank Aggregation}. In \bibinfo{booktitle}{\emph{Proceedings of the 2003 ACM SIGMOD International Conference on Management of Data}}. \bibinfo{pages}{301--312}.
\newblock
\href{https://doi.org/10.1145/872757.872795}{doi:\nolinkurl{10.1145/872757.872795}}


\bibitem[Fagin et~al\mbox{.}(2003b)]%
        {fagin2003optimal}
\bibfield{author}{\bibinfo{person}{Ronald Fagin}, \bibinfo{person}{Amnon Lotem}, {and} \bibinfo{person}{Moni Naor}.} \bibinfo{year}{2003}\natexlab{b}.
\newblock \showarticletitle{Optimal Aggregation Algorithms for Middleware}.
\newblock \bibinfo{journal}{\emph{J. Comput. System Sci.}} \bibinfo{volume}{66}, \bibinfo{number}{4} (\bibinfo{year}{2003}), \bibinfo{pages}{614--656}.
\newblock
\href{https://doi.org/10.1016/S0022-0000(03)00026-6}{doi:\nolinkurl{10.1016/S0022-0000(03)00026-6}}


\bibitem[Hjaltason and Samet(1999)]%
        {hjaltason1999distance}
\bibfield{author}{\bibinfo{person}{G{\'i}sli~R. Hjaltason} {and} \bibinfo{person}{Hanan Samet}.} \bibinfo{year}{1999}\natexlab{}.
\newblock \showarticletitle{Distance Browsing in Spatial Databases}.
\newblock \bibinfo{journal}{\emph{ACM Transactions on Database Systems}} \bibinfo{volume}{24}, \bibinfo{number}{2} (\bibinfo{year}{1999}), \bibinfo{pages}{265--318}.
\newblock
\href{https://doi.org/10.1145/320248.320255}{doi:\nolinkurl{10.1145/320248.320255}}


\bibitem[Li et~al\mbox{.}(2017)]%
        {li2017fexipro}
\bibfield{author}{\bibinfo{person}{Hui Li}, \bibinfo{person}{Tsz~Nam Chan}, \bibinfo{person}{Man~Lung Yiu}, {and} \bibinfo{person}{Nikos Mamoulis}.} \bibinfo{year}{2017}\natexlab{}.
\newblock \showarticletitle{{FEXIPRO}: Fast and Exact Inner Product Retrieval in Recommender Systems}. In \bibinfo{booktitle}{\emph{Proceedings of the 2017 ACM International Conference on Management of Data}}. \bibinfo{pages}{835--850}.
\newblock
\href{https://doi.org/10.1145/3035918.3064009}{doi:\nolinkurl{10.1145/3035918.3064009}}


\bibitem[Lin et~al\mbox{.}(2021)]%
        {lin2021pyserini}
\bibfield{author}{\bibinfo{person}{Jimmy Lin}, \bibinfo{person}{Xueguang Ma}, \bibinfo{person}{Sheng-Chieh Lin}, \bibinfo{person}{Jheng-Hong Yang}, \bibinfo{person}{Ronak Pradeep}, {and} \bibinfo{person}{Rodrigo~Frassetto Nogueira}.} \bibinfo{year}{2021}\natexlab{}.
\newblock \showarticletitle{Pyserini: A Python Toolkit for Reproducible Information Retrieval Research with Sparse and Dense Representations}. In \bibinfo{booktitle}{\emph{Proceedings of the 44th International ACM SIGIR Conference on Research and Development in Information Retrieval}}. \bibinfo{publisher}{ACM}, \bibinfo{pages}{2356--2362}.
\newblock
\href{https://doi.org/10.1145/3404835.3463238}{doi:\nolinkurl{10.1145/3404835.3463238}}


\bibitem[Louis et~al\mbox{.}(2025)]%
        {louis2025know}
\bibfield{author}{\bibinfo{person}{Antoine Louis}, \bibinfo{person}{Gijs van Dijck}, {and} \bibinfo{person}{Gerasimos Spanakis}.} \bibinfo{year}{2025}\natexlab{}.
\newblock \showarticletitle{Know When to Fuse: Investigating {Non-English} Hybrid Retrieval in the Legal Domain}. In \bibinfo{booktitle}{\emph{Proceedings of the 31st International Conference on Computational Linguistics}}. \bibinfo{publisher}{Association for Computational Linguistics}, \bibinfo{address}{Abu Dhabi, UAE}, \bibinfo{pages}{4293--4312}.
\newblock
\urldef\tempurl%
\url{https://aclanthology.org/2025.coling-main.290/}
\showURL{%
\tempurl}


\bibitem[Mallia et~al\mbox{.}(2024)]%
        {mallia2024bmp}
\bibfield{author}{\bibinfo{person}{Antonio Mallia}, \bibinfo{person}{Torsten Suel}, {and} \bibinfo{person}{Nicola Tonellotto}.} \bibinfo{year}{2024}\natexlab{}.
\newblock \showarticletitle{Faster Learned Sparse Retrieval with Block-Max Pruning}. In \bibinfo{booktitle}{\emph{Proceedings of the 47th International ACM SIGIR Conference on Research and Development in Information Retrieval}}. \bibinfo{pages}{2411--2415}.
\newblock
\href{https://doi.org/10.1145/3626772.3657906}{doi:\nolinkurl{10.1145/3626772.3657906}}


\bibitem[Manmatha et~al\mbox{.}(2001)]%
        {manmatha2001score}
\bibfield{author}{\bibinfo{person}{R. Manmatha}, \bibinfo{person}{Toni~M. Rath}, {and} \bibinfo{person}{Fangfang Feng}.} \bibinfo{year}{2001}\natexlab{}.
\newblock \showarticletitle{Modeling Score Distributions for Combining the Outputs of Search Engines}. In \bibinfo{booktitle}{\emph{Proceedings of the 24th Annual International ACM SIGIR Conference on Research and Development in Information Retrieval}}. \bibinfo{pages}{267--275}.
\newblock
\href{https://doi.org/10.1145/383952.384005}{doi:\nolinkurl{10.1145/383952.384005}}


\bibitem[{Microsoft}({[n.\,d.]})]%
        {azure_rrf_docs}
\bibfield{author}{\bibinfo{person}{{Microsoft}}.} \bibinfo{year}{[n.\,d.]}\natexlab{}.
\newblock \bibinfo{title}{Relevance Scoring in Hybrid Search Using Reciprocal Rank Fusion ({RRF})}.
\newblock \bibinfo{howpublished}{\url{https://learn.microsoft.com/en-us/azure/search/hybrid-search-ranking}}.
\newblock
\shownote{Accessed 2026-07-29}.
\newblock


\bibitem[{National Institute of Standards and Technology}(2020a)]%
        {nist_trec_covid_data}
\bibfield{author}{\bibinfo{person}{{National Institute of Standards and Technology}}.} \bibinfo{year}{2020}\natexlab{a}.
\newblock \bibinfo{title}{{TREC-COVID} Data and Evaluation Resources}.
\newblock \bibinfo{howpublished}{\url{https://ir.nist.gov/covidSubmit/data.html}}.
\newblock
\shownote{Accessed 2026-08-04}.
\newblock


\bibitem[{National Institute of Standards and Technology}(2020b)]%
        {nist_trec_covid_round5}
\bibfield{author}{\bibinfo{person}{{National Institute of Standards and Technology}}.} \bibinfo{year}{2020}\natexlab{b}.
\newblock \bibinfo{title}{{TREC-COVID} Round 5 Guidelines}.
\newblock \bibinfo{howpublished}{\url{https://ir.nist.gov/trec-covid/round5.html}}.
\newblock
\shownote{Accessed 2026-08-04}.
\newblock


\bibitem[Roberts et~al\mbox{.}(2021)]%
        {roberts2021treccovid}
\bibfield{author}{\bibinfo{person}{Kirk Roberts}, \bibinfo{person}{Tasmeer Alam}, \bibinfo{person}{Steven Bedrick}, \bibinfo{person}{Dina Demner-Fushman}, \bibinfo{person}{Kyle Lo}, \bibinfo{person}{Ian Soboroff}, \bibinfo{person}{Ellen Voorhees}, \bibinfo{person}{Lucy~Lu Wang}, {and} \bibinfo{person}{William~R. Hersh}.} \bibinfo{year}{2021}\natexlab{}.
\newblock \showarticletitle{Searching for Scientific Evidence in a Pandemic: An Overview of {TREC-COVID}}.
\newblock \bibinfo{journal}{\emph{Journal of Biomedical Informatics}}  \bibinfo{volume}{121} (\bibinfo{year}{2021}), \bibinfo{pages}{103865}.
\newblock
\href{https://doi.org/10.1016/j.jbi.2021.103865}{doi:\nolinkurl{10.1016/j.jbi.2021.103865}}


\bibitem[Robertson and Zaragoza(2009)]%
        {robertson2009bm25}
\bibfield{author}{\bibinfo{person}{Stephen Robertson} {and} \bibinfo{person}{Hugo Zaragoza}.} \bibinfo{year}{2009}\natexlab{}.
\newblock \showarticletitle{The Probabilistic Relevance Framework: {BM25} and Beyond}.
\newblock \bibinfo{journal}{\emph{Foundations and Trends in Information Retrieval}} \bibinfo{volume}{3}, \bibinfo{number}{4} (\bibinfo{year}{2009}), \bibinfo{pages}{333--389}.
\newblock
\href{https://doi.org/10.1561/1500000019}{doi:\nolinkurl{10.1561/1500000019}}


\bibitem[Teflioudi and Gemulla(2016)]%
        {teflioudi2016lemp}
\bibfield{author}{\bibinfo{person}{Christina Teflioudi} {and} \bibinfo{person}{Rainer Gemulla}.} \bibinfo{year}{2016}\natexlab{}.
\newblock \showarticletitle{Exact and Approximate Maximum Inner Product Search with {LEMP}}.
\newblock \bibinfo{journal}{\emph{ACM Transactions on Database Systems}} \bibinfo{volume}{42}, \bibinfo{number}{1} (\bibinfo{year}{2016}), \bibinfo{pages}{1--49}.
\newblock
\href{https://doi.org/10.1145/2996452}{doi:\nolinkurl{10.1145/2996452}}


\bibitem[Thakur et~al\mbox{.}(2021)]%
        {thakur2021beir}
\bibfield{author}{\bibinfo{person}{Nandan Thakur}, \bibinfo{person}{Nils Reimers}, \bibinfo{person}{Andreas R{\"u}ckl{\'e}}, \bibinfo{person}{Abhishek Srivastava}, {and} \bibinfo{person}{Iryna Gurevych}.} \bibinfo{year}{2021}\natexlab{}.
\newblock \showarticletitle{{BEIR}: A Heterogenous Benchmark for Zero-shot Evaluation of Information Retrieval Models}. In \bibinfo{booktitle}{\emph{Proceedings of the Neural Information Processing Systems Track on Datasets and Benchmarks}}, Vol.~\bibinfo{volume}{1}.
\newblock
\showeprint[arxiv]{2104.08663}


\bibitem[Wadden et~al\mbox{.}(2020)]%
        {wadden2020scifact}
\bibfield{author}{\bibinfo{person}{David Wadden}, \bibinfo{person}{Shanchuan Lin}, \bibinfo{person}{Kyle Lo}, \bibinfo{person}{Lucy~Lu Wang}, \bibinfo{person}{Madeleine van Zuylen}, \bibinfo{person}{Arman Cohan}, {and} \bibinfo{person}{Hannaneh Hajishirzi}.} \bibinfo{year}{2020}\natexlab{}.
\newblock \showarticletitle{Fact or Fiction: Verifying Scientific Claims}. In \bibinfo{booktitle}{\emph{Proceedings of the 2020 Conference on Empirical Methods in Natural Language Processing}}. \bibinfo{publisher}{Association for Computational Linguistics}, \bibinfo{pages}{7534--7550}.
\newblock
\href{https://doi.org/10.18653/v1/2020.emnlp-main.609}{doi:\nolinkurl{10.18653/v1/2020.emnlp-main.609}}


\bibitem[Xiao et~al\mbox{.}(2024)]%
        {xiao2024cpack}
\bibfield{author}{\bibinfo{person}{Shitao Xiao}, \bibinfo{person}{Zheng Liu}, \bibinfo{person}{Peitian Zhang}, \bibinfo{person}{Niklas Muennighoff}, \bibinfo{person}{Defu Lian}, {and} \bibinfo{person}{Jian-Yun Nie}.} \bibinfo{year}{2024}\natexlab{}.
\newblock \showarticletitle{{C-Pack}: Packed Resources for General Chinese Embeddings}. In \bibinfo{booktitle}{\emph{Proceedings of the 47th International ACM SIGIR Conference on Research and Development in Information Retrieval}}. \bibinfo{pages}{641--649}.
\newblock
\showeprint[arxiv]{2309.07597}
\href{https://doi.org/10.1145/3626772.3657878}{doi:\nolinkurl{10.1145/3626772.3657878}}


\end{thebibliography}

\appendix

\section{Local Soundness of PVS and PBM Bounds}\label{local-soundness-of-pvs-and-pbm-bounds}

\subsection{Finite-precision guard for PVS}\label{finite-precision-guard-for-pvs}

Section 2.3.1 denotes the quantized center score by \(c(q,v)\). Let \(E(q,v)\) denote the sum of the two quantization-residual bounds given there, and define

\[
\varepsilon=\texttt{f32::EPSILON},
\qquad
a_m=(8m+64)\varepsilon,
\]

with the requirement \(a_m<1\). We assume that certificate construction and authoritative scoring in the frozen implementation satisfy the standard floating-point error model and use at most \(8m+64\) rounded operations. The implementation computes \(\gamma_m\) with an upward-rounded numerator and a downward-rounded denominator so that

\[
\gamma_m\ge \frac{a_m}{1-a_m}.
\]

The finite-precision guard is

\[
G(q,v)=
\left(
|c(q,v)|+E(q,v)+
\|q\|_{2,\uparrow}\|v\|_{2,\uparrow}
\right)\gamma_m.
\]

Assume also that all inputs and intermediate values are finite and that the signed Int8 dot product does not overflow its integer accumulator. Under these assumptions, standard floating-point error accumulation gives

\[
\left|
s_{\mathrm{f32}}(q,v)-c(q,v)
\right|
\le E(q,v)+G(q,v).
\]

Here, \(E(q,v)\) covers the quantization residual, whereas \(G(q,v)\) covers rounding in certificate construction and authoritative scoring. Under these assumptions, rounding the endpoints \(c(q,v)\pm(E(q,v)+G(q,v))\) outward yields an interval containing the authoritative \texttt{f32} score. The PVS exact-prefix guarantee is conditional on these numerical assumptions. PVS fails the request if an input is non-finite, integer accumulation may overflow, or an evaluated authoritative score lies outside its interval.

\subsection{Floating-point range bound for PBM}\label{floating-point-range-bound-for-pbm}

The implementation evaluates both \(S(x)\) and \(H(R)\) from Section 2.3.2 in \texttt{f32}. Missing terms have \(p_t(x)=0\). Query weights, impacts, and block-max metadata must be finite and nonnegative, and the metadata must satisfy

\[
m_t(R)\ge p_t(x),
\qquad
\forall x\in R,\ t\in Q.
\]

Suppose that document scores and range bounds are accumulated in the same fixed term order \(\sigma\). Their machine values are

\[
\widehat S(x)=
\operatorname{fl}_{\sigma}
\left(
\sum_{t\in Q}q_t p_t(x)
\right),
\]

and

\[
\widehat H(R)=
\operatorname{fl}_{\sigma}
\left(
\sum_{t\in Q}q_t m_t(R)
\right).
\]

For every term, the metadata condition and \(q_t\ge0\) bound the document contribution by the corresponding range contribution. Finite, nonnegative floating-point multiplication and addition are monotone. Applying the same operations in the same order therefore preserves the componentwise inequality:

\[
\widehat S(x)\le \widehat H(R),
\qquad
\forall x\in R.
\]

Thus, \(\widehat H(R)\) safely bounds the implementation value \(\widehat S(x)\). PBM fails the request when the block-max metadata does not cover the relevant postings or when the metadata, numerical computation, or read state violates the stated conditions.

\end{document}